%% file: ecocpaper.tex
\documentclass[a4paper, oneside, twocolumn, notitlepage, 10pt]{extarticle_ecoc}
\usepackage{ecoc}

\usepackage[capitalize]{cleveref}

\usepackage[acronym,nomain]{glossaries}
\newcommand{\SetCapsType}{normalcaps}
\input{acronyms.tex}
\begin{document}
\selectlanguage{english}    


\title{Mitigation of Polarization-Induced Fading in Optical Vector Network Analyzer for the Characterization of km-scale Space-Division Multiplexing Fibers}%


\author{
    Besma Kalla \textsuperscript{(1,*)}, Martina Cappelletti \textsuperscript{(1,2)},
    Menno van den Hout \textsuperscript{(1)},
    Vincent van Vliet 
     \textsuperscript{(1)}, \\
    Simon Rommel \textsuperscript{(1)},
    Luca Palmieri\textsuperscript{(2)},
    Thomas Bradley\textsuperscript{(1)} and Chigo Okonkwo\textsuperscript{(1)}
}

\maketitle                  

\begin{strip}
    \begin{author_descr}

        \textsuperscript{(1)} High Capacity Optical Transmission Laboratory, Eindhoven University of Technology, the Netherlands,

        \textsuperscript{(2)} Department of Information Engineering, University of Padova, Padova, Italy, \textcolor{blue}{\uline{*b.kalla@tue.nl}}

    \end{author_descr}
\end{strip}

\renewcommand\footnotemark{}
\renewcommand\footnoterule{}


\begin{strip}
    \begin{ecoc_abstract}
         We propose an optimized optical vector network analyzer with automatic polarization control to stabilize the reference arm polarization throughout the sweep range. We demonstrate this technique, successfully removing the polarization-induced fading and measurement distortions in insertion loss by characterizing a 10~km multi-core fiber.
          \textcopyright\,2024 The Author(s)
    \end{ecoc_abstract}
\end{strip}
\glsresetall

\section{Introduction}
\begin{figure*}[b]
    \centering
    \includegraphics[width=\linewidth]{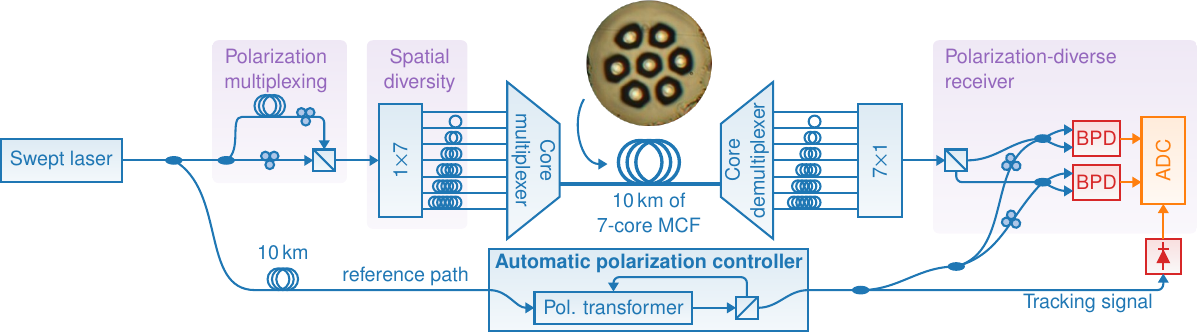}
    \caption{SDM-OVNA setup optimized with an \lAPC and configured to measure 7-core multi-core fiber}
    \label{fig:setup}
\end{figure*}

\uSDM has proven to be a promising technology \cite{puttnam_space-division_2021,puttnam_229_2023} to increase capacity beyond that of single-mode fiber for long-haul transmission, as it exploits multiple spatial channels for data transmission. Various SDM fiber architectures such as \FMF, uncoupled and coupled \MCF, and \FMMCF have been widely characterized in the laboratory environment. More recently, deployed fibers are increasingly being studied\cite{rademacher_characterization_2022}. In these experiments, the performance of optical fibers in spooled, unspooled, or twisted conditions has been shown to present different optical performances depending on the coupling regime \cite{mazur_transfer_2020}. The same type of fiber also showed distinct characteristics \cite{mazur_broadband_2023}, highlighting the unique impact of various deployment scenarios.

For the practical implementation of \SDM technology, it is important to characterize fibers at every stage, from spooled to cabled and finally deployed. An \OVNA is a powerful characterization tool in this context. It enables the measurement of the full linear transfer function matrix of \SDM systems in a single scan, comprehensively characterizing the optical properties.
Based on \lSWI, \cite{vanwiggeren_swept-wavelength_2003} an \OVNA utilizes one branch of the interferometer for the device or fiber-under-test, and the other is a reference fiber. The reference path must be length matched to that of the \DUT to maintain coherence between both arms to ensure accurate phase and amplitude measurements across the sweep range of the laser. 

However, a challenge arises when measuring kilometer-scale fibers. For the OVNA, the signal in the reference arm undergoes wavelength-dependent polarization rotations, leading to a wavelength-dependent power fading in the beating signal at the receiver end. This results in measurement distortions of the linear fiber-under-test parameters. Previous works demonstrated a digital polarization equalization technique that improved the measurements \cite{rommel_polarization_2019}, but equalization becomes impossible in deep-fading scenarios. Another method includes adding a Faraday mirror in the reference arm \cite{mazur_transfer_2020}, with the reference fiber being half the length of the \DUT \cite{mazur_broadband_2023}. However,  the effectiveness of this approach for OVNA measurements has not yet been reported.


%


In this paper, we present an optimized OVNA setup to mitigate polarization-induced fading using an \APC placed in the reference arm, which allows the characterization of kilometer-scale fiber links, representative of real-deployment scenarios, without additional requirements on the transmitter or receiver hardware. We demonstrate the effectiveness of this configuration by characterizing a 10~km long \MCF with seven uncoupled cores. Our results indicate a more accurate measurement of the \IL derived from the transfer function with automatic polarization control.

\section{Principle of Operation}

The \OVNA, shown in \cref{fig:setup}, consists of a \STL fed into an interferometric structure with a measurement and reference branch. The non-linear frequency sweep of the laser is corrected for with an auxiliary interferometer\cite{rommel_few-mode_2017}, not shown in the diagram. A polarization-multiplexing stage at the input allows the launching of the two time-delayed polarizations of the optical field. \uPolCs are used to obtain two orthogonal and equal-power polarizations. Additionally, the setup includes power splitters with different delay lines at the input and output of the \DUT for \SDM device measurements \cite{fontaine_characterization_2013}. In the polarization-diverse receiver, the signal is split into two orthogonal polarizations, each interfered with the polarization-aligned reference to detect both. The two output signals X and Y are then digitized using an \ADC.

To obtain maximum beating between the measurement X and Y polarization path and the reference signal, polarization alignment is needed. This is done by maximizing the beating power using \PolCs in the reference path. Assuming the state of polarization in the reference arm remains constant over the full sweep range of the laser, this polarization alignment can be achieved by maximizing the power of the fringe pattern. Finally, the complex transfer function is retrieved by following a sequence of \DSP steps \cite{rommel_few-mode_2017}, from which the linear device parameters, such as \IL, \XT, and \MDL, can be calculated. 

The reference arm of the \OVNA contains an optical delay line to ensure coherence between the measurement and reference arms, by approximately matching their path lengths. Moreover, the path length difference should be kept minimal, as the frequency of the fringe pattern is the product of the sweeping rate and this delay. Hence, a large path difference results in higher frequency, requiring higher speed digitizers.  
This means that to characterize km-scale SDM fibers, a reference of similar length is required. 

A challenge arises as the state of polarization of the signal propagating through the long reference fiber delay changes across the sweep due to wavelength-dependent birefringence in the fiber. Consequently, the varying polarization leads to wavelength-dependent fading as the reference polarization is no longer matched to the signal polarization in the polarization diverse receiver. Due to this fading, the measurements of the linear device parameters are distorted.

Hence, we propose adding an \APC feature with an optical feedback loop in the reference arm of the \OVNA, as shown in \cref{fig:setup}, to control the polarization across the entire wavelength sweep range. Using a LiNbO\textsubscript{3}-based polarization transformer \cite{noe_linbo3-based_2016}, the \APC continuously aligns the state of polarization of the reference signal to the expected state for the polarization diverse receiver by minimizing power in the orthogonal polarization. This results in a constant polarization state throughout the measurement sweep, meaning that aligning the reference and the signal at one wavelength ensures co-polarized signals throughout the measurement sweep, thus eliminating polarization-induced fading. A small fraction of the light in the reference at the output of the \APC is used for tracking power, referred to as tracking signal. As the light is polarized, any variation in the measured power throughout the measurement sweep is indicative of polarization rotation.

Note that the polarization control has to be fast enough to keep track of the polarization rotation occurring in the reference arm. The variation of the state of polarization can be estimated by the polarization rotation rate R, which is calculated as 
\begin{equation}
   R = 2 \pi \gamma T,
\end{equation}
where $\gamma$ is the frequency sweep rate of the \STL in Hz/s and $T$ the differential group delay of the reference fiber.

\section{Fading-free Fiber Insertion Loss Measurement}

We employ the proposed polarization tracking and control method to characterize a 10~km \MCF as proof of principle with the OVNA configured as shown in \cref{fig:setup} using spatial multiplexers based on laser-inscribed glass waveguides \cite{gross_ultrafast-laser-inscribed_2015}. A key parameter that describes SDM device properties is \IL, defined as the average of the squared singular values of the complex transfer matrix\cite{rommel_few-mode_2017}. We analyze \IL for this \MCF to highlight the measurement distortions caused by the polarization rotation when measuring with fibers of km-scale lengths. The laser is swept at 100~nm/s from 1530~nm to 1570~nm. The tracking speed required for this configuration is estimated to be approximately 50~rad/s. The transfer function is extracted digitally from which the \IL is calculated. As the focus of this study is the wavelength-dependent variation in the system, we normalize the tracking signal at the output of the \APC and the \IL to the optimized values obtained at single-wavelength calibration.
\begin{figure}[h!]
    \centering
    \includegraphics[width=\linewidth]{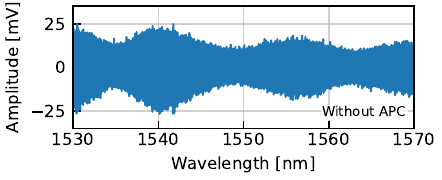}
    \includegraphics[width=\linewidth]{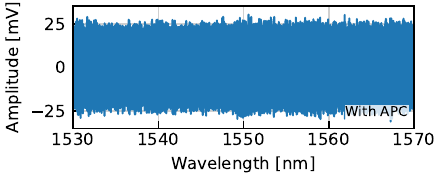}
    \caption{Digitized waveform of one receiver output channel: (top) without (bottom) with automatic polarization control. }
    \label{fig:wavefrom}
\end{figure}

\begin{figure}[h!]
    \vspace{2mm}
    \centering
    \includegraphics[width=\linewidth]{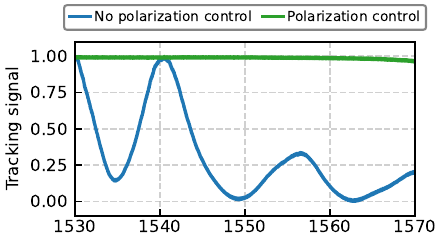}
    \includegraphics[width=\linewidth]{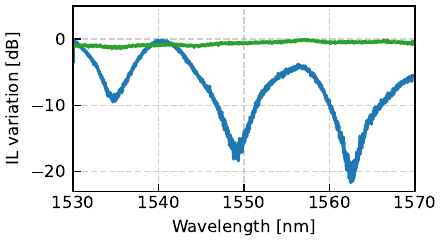}
    \caption{(Top) Normalized tracking signal of the reference arm signal. (Bottom) Normalized insertion loss of the central core of the \MCF }
    \label{fig:comp_plot}
\end{figure}

We first focus on analyzing the results of a single core to demonstrate the effect of polarization rotation. The sampled waveform at one of the polarization-diverse receiver's \BPDs is shown in \cref{fig:wavefrom}. The top figure shows that without automatic polarization control, there is substantial power fluctuation across the measurement wavelength range. The fading is correlated to the trend of the tracking signal, as can be seen at the top of \cref{fig:comp_plot}, confirming that the fading in the fringe pattern is induced by the polarization rotation. The \IL derived from such measurement is distorted as shown in the bottom of \cref{fig:comp_plot}. Significant dips in the \IL are observed in the wavelengths where the fading occurs due to lower reference power. 

 However, with automatic polarization control, the polarization in the reference arm can be stabilized across wavelengths, as seen on the tracking signal at the top of \cref{fig:comp_plot}, meaning that the reference and measurement signals are aligned at all wavelengths. This eliminates the fading as can be seen at the bottom of \cref{fig:wavefrom} and provides more accurate \IL calculation across wavelength as shown at the bottom of \cref{fig:comp_plot}. Note that the slight decrease in power in the tracking signal after 1560~nm is due to the laser variation with wavelength and not from the polarization rotation.

\begin{figure}[h]
    \centering
    \includegraphics[width=\linewidth]{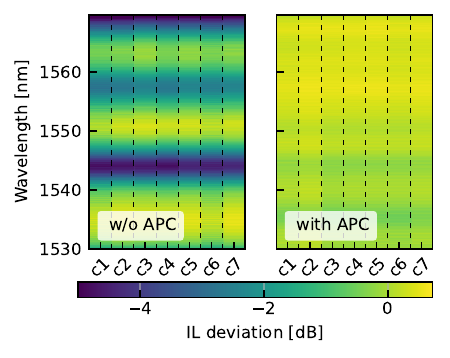}
    \caption{Insertion loss variation with wavelength for the 7 cores of the \MCF without and with automatic polarization control, where Cn is the specific core measured}
    \label{fig:colormap}
\end{figure}

We then proceed to analyze the \IL variation with wavelength for all 7 cores of the \MCF. \Cref{fig:colormap} shows the variation in \IL, relative to the \IL where the system optimization was done (1550~nm), across the measurement sweep with and without automatic polarization control. It can be seen that without \APC, there is a strong deviation from the optimized IL of up to 4~dB as a result of the fading. However, with \APC, the variation is below 0.5~dB with a minor increase in \IL from 1550 to 1570~nm, more characteristic of the true wavelength-dependent properties of the fiber. The standard deviation of the wavelength dependent \IL of the MCF is reduced from 1.49 to 0.35~dB.

\section{Conclusions} We have proposed an optimized \OVNA setup with automatic polarization control to mitigate the impact of polarization rotation in the reference arm and ensure proper alignment between the reference and the signal in the polarization-diverse receiver, eliminating the polarization-induced power fading. We demonstrate that this improves the wavelength-dependent \IL measured with the \OVNA for a 10~km 7-core \MCF.

\clearpage
\section{Acknowledgements}
We thank Novoptel and Professor Reinhold Noé and his team for support with the automatic polarization controller. 
This work acknowledges partial funding by the European Union’s Horizon Europe Research and Innovation Programme HOMTech Project under the Marie Sklodowska-Curie Grant Agreement No. 101072409 and from the Dutch Ministry of Economic Affairs and Climate Policy, under the PhotonDelta National GrowthFunds.


\printbibliography[]

\vspace{-4mm}


\end{document}



%% file: acronyms.tex
\usepackage{xspace}  
\usepackage[shortcuts]{extdash}
\usepackage{listofitems,pgffor}    
\usepackage{siunitx}
\usepackage{xstring}    
\usepackage{silence}
\usepackage{xparse}

\setsepchar{;}  

\ifdefined\silencecommonwarnings
\else
	\def\silencecommonwarnings{true} 
\fi

\ifbool{\silencecommonwarnings}{%
    \WarningFilter{ECOtools}{Cannot define: DH}%
    \WarningFilter{ECOtools}{Cannot define: PAM}%
    \WarningFilter{ECOtools}{Cannot define: QAM}%
    \WarningFilter{ECOtools}{Cannot define: SI}%
    \WarningFilter{ECOtools}{Cannot define: PV}%
    \WarningFilter{ECOtools}{Cannot define: LP}%
    \WarningFilter{ECOtools}{Cannot define: uLP}%
    \WarningFilter{ECOtools}{Redefining DH}%
    }{}

\makeatletter
\providecommand{\SetCapsType}{smallcaps}

\long\def\@scTrue{smallcaps}
\long\def\@scFalse{normalcaps}
\newcommand{\acroSCaps}[1]{%
    \ifx\SetCapsType\@scTrue 
        \textsc{#1}%
    \else
        \MakeUppercase{#1}%
    \fi
}
\makeatother

\usepackage{scalerel}
\makeatletter
\newcommand\scslash{%
\ifx\SetCapsType\@scTrue 
    \protect\stretchrel*{$/$}{\textsc{e}}
\else
    /
\fi
} 
\makeatother 

\makeatletter
\@ifpackageloaded{babel}{%
    \newcommand{\usuk}[2]{%
        \iflanguage{USenglish}{#1}{#2}%
    }%
}{%
    \newcommand{\usuk}[2]{%
        #1%
    }%
}%

\newcommand{\langcheck}[2]{
    \@ifpackageloaded{babel}{%
        \iflanguage{USenglish}{#1}{#2}%
    }{%
        #1%
    }%
}

\makeatother

\newcommand{\short}[1]{%
    \glsentrytext{#1}\xspace%
}
\newcommand{\shortfakeplural}[1]{%
    \glsentrytext{#1}s\xspace%
}
\newcommand{\Short}[1]{%
    \Glsentrytext{#1}\xspace%
}
\newcommand{\normal}[1]{%
    \gls{#1}\xspace%
}
\newcommand{\longacr}[1]{%
    \acrlong{#1}\xspace%
}
\newcommand{\plural}[1]{%
    \glspl{#1}\xspace%
}
\newcommand{\full}[1]{%
    \acrfull{#1}\xspace%
}
\newcommand{\fullplural}[1]{%
    \acrfullpl{#1}\xspace%
}
\newcommand{\Normal}[1]{%
    \Gls{#1}\xspace%
}
\newcommand{\Plural}[1]{%
    \Glspl{#1}\xspace%
}
\newcommand{\Full}[1]{%
    \Acrfull{#1}\xspace%
}
\newcommand{\Fullplural}[1]{%
    \Acrfullpl{#1}\xspace%
} 

\newcommand{\texpdfif}[2]{%
    \ifcsname texorpdfstring\endcsname%
        \texorpdfstring{#1{#2}}{#2\xspace}%
    \else%
        #1{#2}%
    \fi%
}

\newcommand{\checkanddefine}[3]{%
	\ifcsname #1\endcsname%
        \PackageWarning{ECOtools}{Cannot define: #1 already defined, trying to define g#1 instead.}%
        \ifcsname g#1\endcsname%
            \PackageWarning{ECOtools}{Cannot define: g#1 also already defined.}%
    	\else%
        	\expandafter\newcommand\csname g#1\endcsname{%
        	    \texpdfif{#2}{#3}%
    	    }%
        \fi%
	\else%
    	\expandafter\newcommand\csname #1\endcsname{%
    	    \texpdfif{#2}{#3}%
	    }%
    \fi%
}

\newcommand{\redefine}[3]{%
    \PackageWarning{ECOtools}{Redefining #1}%
	\expandafter\renewcommand\csname #1\endcsname{%
	    \texpdfif{#2}{#3}%
    }%
}

\newcommand{\nAcronym}[4][]{%
	\newacronym[#1]{#2}{#3}{#4}%
	\checkanddefine{s#2}{\short}{#2}%
    \checkanddefine{s#2s}{\shortfakeplural}{#2}%
	\checkanddefine{#2}{\normal}{#2}%
	\checkanddefine{l#2}{\longacr}{#2}%
	\checkanddefine{#2s}{\plural}{#2}%
	\checkanddefine{f#2}{\full}{#2}%
	\checkanddefine{f#2s}{\fullplural}{#2}%
	\checkanddefine{su#2}{\Short}{#2}%
	\checkanddefine{u#2}{\Normal}{#2}%
	\checkanddefine{u#2s}{\Plural}{#2}%
	\checkanddefine{fu#2}{\Full}{#2}%
	\checkanddefine{fu#2s}{\Fullplural}{#2}%
	\IfStrEq{#2}{DH}{
	    \redefine{#2}{\normal}{#2}%
	    }{}%
}%

\NewDocumentCommand\qam{g}{%
    \IfNoValueTF{#1}{%
        \texpdfif{\gls}{QAM}\xspace%
        }{%
        \StrLen{#1}[\stringlength]%
        \ifnum\stringlength=0%
            \texpdfif{\gls}{QAM}\xspace%
        \else%
            {\qamlisthelper{#1}}%
        \fi%
        }%
}

\let\QAM\qam

\DeclareRobustCommand\qamlisthelper[1]{%
    \readlist*\args{#1}%
    \acroSCaps{\args[1]\=/}%
    \ifnum\argslen = 2%
        { and \acroSCaps{\args[2]}\=/}%
    \fi%
    \ifnum\argslen > 2%
        \foreach \n in {2,...,\argslen}{%
            \ifnum\n = \argslen%
                {, and }%
            \else 
                {, }%
            \fi%
            {\acroSCaps{\args[\n]}\=/}%
        }%
    \fi%
    \ifglsused{QAM}%
        {}%
        {ary }%
    \texpdfif{\gls}{QAM}%
}%

\NewDocumentCommand\pam{g}{%
    \IfNoValueTF{#1}{%
        \texpdfif{\gls}{PAM}\xspace%
        }{%
        \StrLen{#1}[\stringlength]%
        \ifnum\stringlength=0%
            \texpdfif{\gls}{PAM}\xspace%
        \else%
            {\pamlisthelper{#1}}%
        \fi%
        }%
}

\DeclareRobustCommand\pamlisthelper[1]{%
    \readlist*\args{#1}%
    \ifglsused{PAM}{%
        \texpdfif{\gls}{PAM}%
        \acroSCaps{\=/\args[1]}%
        \ifnum\argslen = 2%
            { and \=/\acroSCaps{\args[2]}}%
        \fi%
        \ifnum\argslen > 2%
            \foreach \n in {2,...,\argslen}{%
                \ifnum\n = \argslen%
                    {, and }%
                \else%
                    {, }%
                \fi%
                {\=/\acroSCaps{\args[\n]}}%
            }%
        \fi%
    }{%
        \acroSCaps{\args[1]\=/}%
        \ifnum\argslen = 2%
            { and \acroSCaps{\args[2]}\=/}%
        \fi%
        \ifnum\argslen > 2%
            \foreach \n in {2,...,\argslen}{%
                \ifnum\n = \argslen%
                    {, and }%
                \else%
                    {, }%
                \fi
                {\acroSCaps{\args[\n]}\=/}%
            }%
        \fi%
        {ary }%
        \texpdfif{\gls}{PAM}%
    }%
}%

\NewDocumentCommand\lp{g}{%
    \IfNoValueTF{#1}{%
        \texpdfif{\normal}{LP}%
        }{%
        \StrLen{#1}[\stringlength]%
        \ifnum\stringlength=0%
            \texpdfif{\normal}{LP}%
        \else%
            \ifglsused{LP}{}{\texpdfif{\normal}{LP}\xspace}%
            \lplisthelper[lp]{#1}%
        \fi%
        }%
}

\NewDocumentCommand\ulp{g}{%
    \IfNoValueTF{#1}{%
        \texpdfif{\Normal}{LP}\xspace%
        }{%
        \StrLen{#1}[\stringlength]%
        \ifnum\stringlength=0%
            \texpdfif{\Normal}{LP}\xspace%
        \else%
            \ifglsused{LP}{%
                \lplisthelper[Lp]{#1}%
            }{%
                \texpdfif{\Normal}{LP}\xspace\lplisthelper[lp]{#1}%
            }%
        \fi%
        }%
}
\DeclareRobustCommand\lplisthelper[2][lp]{%
    \readlist*\args{#2}%
    \foreach \n in {1,...,\argslen}{%
        \ifnum \n > 1%
            \ifnum \argslen > 2%
                {, }%
            \else%
                { }%
            \fi%
        \fi%
        \ifnum \n = \argslen%
            \ifnum \argslen > 1%
                {and }%
            \fi%
        \fi%
        \ifnum \n = 1%
            {\acroSCaps{#1}}
        \else%
            {\acroSCaps{\MakeLowercase{#1}}}%
        \fi%
        {\textsubscript{\StrSplit{\args[\n]}{2}{\csA}{\csB}\acroSCaps{\csA}\csB}}
    }%
}%

\nAcronym{128SPQAM}{\acroSCaps{128-sp-16-qam}}{128-ary set-partitioning \QAM{16}}

\nAcronym{2A8PSK}{\acroSCaps{2a8psk}}{2-ary amplitude 8-ary phaseshift keying}

\nAcronym{3CCMCF}{\acroSCaps{3cc-mcf}}{3-core coupled-core multi-core fiber}

\nAcronym{4D}{\acroSCaps{4d}}{four-dimensional}
\nAcronym{4D64PRS}{\acroSCaps{4d-64prs}}{\usuk{four-dimensional 64-ary polarization-ring-switching}{four-dimensional 64-ary polarisation-ring-switching}}
\nAcronym{4DOS128}{\acroSCaps{4d-os128}}{four-dimensional orthant-symmetric 128-ary modulation format}

\nAcronym{5B4D2A8PSK}{\acroSCaps{5b4d-2a8psk}}{5-bit four-dimensional two-amplitude 8-ary phase-shift keying}

\nAcronym{6B4D2A8PSK}{\acroSCaps{6b4d-2a8psk}}{6-bit four-dimensional two-amplitude 8-ary phase-shift keying}

\nAcronym{7B4D2A8PSK}{\acroSCaps{7b4d-2a8psk}}{7-bit four-dimensional two-amplitude 8-ary phase-shift keying}

\nAcronym{8D}{\acroSCaps{8d}}{eight-dimensional}\nAcronym{8D2048PRS}{\acroSCaps{8d-2048prs}}{eight-dimensional 2048-ary polarization-ring-switching}
\nAcronym{8D2048PRST1}{\acroSCaps{8d-2048prs-t1}}{eight-dimensional 2048-ary polarization-ring-switching type 1}
\nAcronym{8D2048PRST2}{\acroSCaps{8d-2048prs-t2}}{eight-dimensional 2048-ary polarization-ring-switching type 2}
\nAcronym{8DAPSK}{\acroSCaps{8d-apsk}}{eight-dimensional amplitude-phase-shift keying}

\nAcronym{ABC}{\acroSCaps{abc}}{automatic bias controller}
\nAcronym{AC}{\acroSCaps{ac}}{alternating current}
\nAcronym{ADC}{\acroSCaps{adc}}{analog-to-digital converter}
\nAcronym{AGC}{\acroSCaps{agc}}{automatic gain control}
\nAcronym{AIR}{\acroSCaps{air}}{achievable information rate}
\nAcronym{AO}{\acroSCaps{ao}}{adaptive optics}
\nAcronym{AOM}{\acroSCaps{aom}}{acousto-optic modulator}
\nAcronym{APC}{\acroSCaps{apc}}{\usuk{automatic polarization controller}{automatic polarisation controller}}
\nAcronym{APD}{\acroSCaps{apd}}{avalanche photodiode}
\nAcronym{API}{\acroSCaps{api}}{application programming interface}
\nAcronym{AR}{\acroSCaps{ar}}{achievable rate}
\nAcronym{ARRWG}{\acroSCaps{a}rr\acroSCaps{wg}}{arrayed-waveguide grating}
\nAcronym{ASE}{\acroSCaps{ase}}{amplified spontaneous emission}
\nAcronym{ASK}{\acroSCaps{ask}}{amplitude-shift keying}
\nAcronym{ASIC}{\acroSCaps{asic}}{application-specific integrated circuit}
\nAcronym{ATS}{\acroSCaps{ats}}{alignment tracking sensor}
\nAcronym{AWG}{\acroSCaps{awg}}{arbitrary-waveform generator}
\nAcronym{AWGN}{\acroSCaps{awgn}}{additive white Gaussian noise}

\nAcronym{BBU}{\acroSCaps{bbu}}{baseband unit}
\nAcronym{BCH}{\acroSCaps{bch}}{Bose-Chaudhuri-Hocquenghem}
\nAcronym{BER}{\acroSCaps{ber}}{bit error rate}
\nAcronym{BERT}{\acroSCaps{bert}}{bit error rate tester}
\nAcronym{BICM}{\acroSCaps{bicm}}{bit-interleaved coded modulation}
\nAcronym{BMD}{\acroSCaps{bmd}}{bit-metric decoding}
\nAcronym{BPD}{\acroSCaps{bpd}}{balanced photo-diode}
\nAcronym{BPF}{\acroSCaps{bpf}}{bandpass filter}
\nAcronym{BPS}{\acroSCaps{bps}}{blind phase search}
\nAcronym{BPSK}{\acroSCaps{bpsk}}{binary phase-shift keying}
\nAcronym{BRGC}{\acroSCaps{brgc}}{binary reflected Gray code}
\nAcronym{BTB}{\acroSCaps{btb}}{back-to-back}

\nAcronym{CAGR}{\acroSCaps{cagr}}{compound annual growth rate}
\nAcronym{CCDM}{\acroSCaps{ccdm}}{constant composition distribution matching}
\langcheck{%
    \nAcronym{CCF}{\acroSCaps{ccf}}{coupled-core fiber}%
    }{%
    \nAcronym{CCF}{\acroSCaps{ccf}}{coupled-core fibre}%
}%
\nAcronym{CD}{\acroSCaps{cd}}{chromatic dispersion}
\nAcronym{CIR}{\acroSCaps{cir}}{channel impulse response}
\nAcronym{CMA}{\acroSCaps{cma}}{constant modulus algorithm}
\nAcronym{CMF}{\acroSCaps{cmf}}{core multiplicity factor}
\nAcronym{CMUX}{\acroSCaps{cmux}}{core multiplexer}
\nAcronym{COTS}{\acroSCaps{cots}}{commercial off-the-shelf}
\nAcronym{ChUT}{\acroSCaps{chut}}{channel under test}
\nAcronym[firstplural=channels under test (\acroSCaps{cut}s)]{CUT}{\acroSCaps{cut}}{channel under test}
\nAcronym{CRX}{\acroSCaps{crx}}{coherent receiver}
\nAcronym{CPE}{\acroSCaps{cpe}}{carrier phase estimation}
\nAcronym{CPU}{\acroSCaps{cpu}}{central processing unit}
\nAcronym{CSPR}{\acroSCaps{cspr}}{carrier-to-signal power ratio}
\nAcronym{CUDA}{\acroSCaps{cuda}}{compute unified device architecture}
\nAcronym{CVQKD}{\acroSCaps{cv-qkd}}{continuous-variable quantum key distribution}
\nAcronym{CW}{\acroSCaps{cw}}{continuous wave}
\nAcronym{CCD}{\acroSCaps{ccd}}{charge-coupled device}

\nAcronym{DA}{\acroSCaps{da}}{driver amplifier}
\nAcronym{DAC}{\acroSCaps{dac}}{digital-to-analog converter}
\nAcronym{DC}{\acroSCaps{dc}}{direct current}
\nAcronym{DBP}{\acroSCaps{dbp}}{digital backpropagation}
\nAcronym{DCF}{\acroSCaps{dcf}}{\usuk{dispersion-compensating fiber}{dispersion-compensating fibre}}
\langcheck{%
    \nAcronym{DCI}{\acroSCaps{dci}}{data center interconnect}
    }{%
    \nAcronym{DCI}{\acroSCaps{dci}}{data centre interconnect}
}%
\nAcronym{DDLMS}{\acroSCaps{dd-lms}}{decision-directed least mean square}
\nAcronym{DEMUX}{\acroSCaps{demux}}{de-multiplexer}
\nAcronym{DFA}{\acroSCaps{dfa}}{\usuk{doped fiber amplifier}{doped fibre amplifier}}
\nAcronym{DFB}{\acroSCaps{dfb}}{distributed feedback}
\nAcronym{DGD}{\acroSCaps{dgd}}{differential group delay}
\nAcronym{DH}{\acroSCaps{dh}}{digital holography}
\nAcronym{DM}{\acroSCaps{dm}}{distribution matcher}
\nAcronym{DMR}{\acroSCaps{dm}}{dichroic mirror}
\nAcronym{DMA}{\acroSCaps{dma}}{direct memory access}
\nAcronym{DMD}{\acroSCaps{dmd}}{differential mode delay}
\nAcronym{DMG}{\acroSCaps{dmg}}{differential modal gain}
\nAcronym{DMGD}{\acroSCaps{dmgd}}{differential mode group delay}
\nAcronym{DML}{\acroSCaps{dml}}{directly-modulated laser}
\nAcronym{DP}{\acroSCaps{dp}}{\usuk{dual-polarization}{dual-polarisation}}
\nAcronym{DPC}{\acroSCaps{dpc}}{digital pre-compensation}
\nAcronym{DPE}{\acroSCaps{dpe}}{digital pre-emphasis}
\nAcronym{DPIQ}{\acroSCaps{dp-iqm}}{\usuk{dual-polarization \acroSCaps{iq}-modulator}{dual-polarisation \acroSCaps{iq}-modulator}}
\nAcronym{DPLL}{\acroSCaps{dpll}}{digital phase-locked loop}
\nAcronym{DQPSK}{\acroSCaps{dqpsk}}{differential quaternary phase-shift-keying}
\nAcronym{DRA}{\acroSCaps{dra}}{distributed Raman amplifier}
\nAcronym{DRE}{\acroSCaps{dre}}{digital resolution enhancer}
\nAcronym{DSB}{\acroSCaps{dsb}}{double-sideband}
\nAcronym{DSF}{\acroSCaps{dsf}}{\usuk{dispersion-shifted fiber}{dispersion-shifted fibre}} 
\nAcronym{DSO}{\acroSCaps{dso}}{digital sampling oscilloscope}
\nAcronym{DSP}{\acroSCaps{dsp}}{digital signal processing}
\nAcronym{DUT}{\acroSCaps{dut}}{device-under-test}
\nAcronym{DWDM}{\acroSCaps{dwdm}}{dense wavelength-division multiplexing}

\nAcronym{EAM}{\acroSCaps{eam}}{electro-absorption modulator}
\nAcronym{ECL}{\acroSCaps{ecl}}{external cavity laser}
\nAcronym{ED}{\acroSCaps{ed}}{Eucledian distance}
\nAcronym{EDF}{\acroSCaps{edf}}{\usuk{erbium-doped fiber}{erbium-doped fibre}}
\nAcronym{EDFA}{\acroSCaps{edfa}}{\usuk{erbium-doped fiber amplifier}{erbium-doped fibre amplifier}}
\nAcronym{ENOB}{\acroSCaps{enob}}{effective number of bits}
\nAcronym{ESS}{\acroSCaps{ess}}{enumerative sphere shaping}

\langcheck{%
    \nAcronym{FBG}{\acroSCaps{fbg}}{fiber Bragg grating}%
    }{%
    \nAcronym{FBG}{\acroSCaps{fbg}}{fibre Bragg grating}%
}%
\nAcronym{FD}{\acroSCaps{fd}}{frequency domain}
\nAcronym{FDE}{\acroSCaps{fde}}{\usuk{frequency domain equalizer}{frequency domain equaliser}}
\nAcronym{FEC}{\acroSCaps{fec}}{forward error correction}
\nAcronym{FFE}{\acroSCaps{ffe}}{\usuk{feed-forward equalizer}{feed-forward equaliser}}
\nAcronym{FFT}{\acroSCaps{fft}}{fast Fourier transform}
\nAcronym{FIR}{\acroSCaps{fir}}{finite impulse response}
\nAcronym{FLOPS}{\acroSCaps{flops}}{floating point operations per second}
\nAcronym{FMEDF}{\acroSCaps{fm-edf}}{\usuk{few-mode erbium-doped fiber}{few-mode erbium-doped fibre}}
\nAcronym{FMEDFA}{\acroSCaps{fm-edfa}}{\usuk{few-mode erbium-doped fiber amplifier}{few-mode erbium-doped fibre amplifier}}
\langcheck{%
    \nAcronym{FMF}{\acroSCaps{fmf}}{few-mode fiber}%
    }{%
    \nAcronym{FMF}{\acroSCaps{fmf}}{few-mode fibre}%
}%
\nAcronym[plural=FM-MCF, firstplural=\usuk{few-mode multi-core fibers}{few-mode multi-core fibres}]{FMMCF}{\acroSCaps{fm-mcf}}{\usuk{few-mode multi-core fiber}{few-mode multi-core fibre}}
\langcheck{%
    \nAcronym{FMPBGF}{\acroSCaps{fm-pbgf}}{few-mode photonic bandgap fiber}%
    }{%
    \nAcronym{FMPBGF}{\acroSCaps{fm-pbgf}}{few-mode photonic bandgap fibre}%
}%
\nAcronym{FOV}{\acroSCaps{fov}}{field of view}
\nAcronym{FPGA}{\acroSCaps{fpga}}{field-programmable gate array}
\nAcronym{FWM}{\acroSCaps{fwm}}{four-wave mixing}
\nAcronym{FSO}{\acroSCaps{fso}}{free-space optical}
\nAcronym{FUT}{\acroSCaps{fut}}{\usuk{fiber under test}{fibre under test}}

\nAcronym{GBP}{\acroSCaps{gbp}}{gain bandwidth product}
\nAcronym{GD}{\acroSCaps{gd}}{group delay}
\nAcronym{GI}{\acroSCaps{gi}}{graded-index}
\nAcronym{GFF}{\acroSCaps{gff}}{gain flattening filter}
\nAcronym{GIFMF}{\acroSCaps{gi-fmf}}{\usuk{graded-index few-mode fiber}{graded-index few-mode fibre}}
\nAcronym{GIMMF}{\acroSCaps{gi-mmf}}{\usuk{graded-index multi-mode fiber}{graded-index multi-mode fibre}}
\nAcronym{GMI}{\acroSCaps{gmi}}{\usuk{generalized mutual information}{generalised mutual information}}
\nAcronym{GNSE}{\acroSCaps{gnse}}{generalized nonlinear Schr\"{o}dinger equation}
\nAcronym{GPU}{\acroSCaps{gpu}}{graphics processing unit}
\nAcronym{GS}{\acroSCaps{gs}}{geometric shaping}
\nAcronym{GV}{\acroSCaps{gv}}{group velocity}
\nAcronym{GVD}{\acroSCaps{gvd}}{group velocity dispersion}
\nAcronym{GPIO}{\acroSCaps{gpio}}{general purpose input output}
\nAcronym{GUI}{\acroSCaps{gui}}{graphical user interface}

\langcheck{%
    \nAcronym{HCF}{\acroSCaps{hcf}}{hollow-core fiber}
    }{%
    \nAcronym{HCF}{\acroSCaps{hcf}}{hollow-core fibre}
}%
\nAcronym{HDFEC}{\acroSCaps{hd-fec}}{hard-decision forward error correction}
\nAcronym{HG}{\acroSCaps{hg}}{Hermite-Gaussian}
\nAcronym{HOM}{\acroSCaps{hom}}{higher-order modes}
\nAcronym{HV}{\acroSCaps{hv}}{Hufnagel-Valley}
\nAcronym{HAP}{\acroSCaps{hap}}{Hufnagel-Andrew-Phillips}

\nAcronym{ICR}{\acroSCaps{icr}}{integrated coherent receiver}
\nAcronym{ICS}{\acroSCaps{ics}}{inter-core skew}
\nAcronym{ICXT}{\acroSCaps{ic-xt}}{inter-core cross-talk}
\nAcronym[plural=IL, firstplural=insertion losses (\acroSCaps{il})]{IL}{\acroSCaps{il}}{insertion loss}
\nAcronym{IFFT}{\acroSCaps{ifft}}{inverse fast Fourier transform}
\nAcronym{IIR}{\acroSCaps{iir}}{intensity impulse response}
\nAcronym{IM}{\acroSCaps{im}}{intensity modulator}
\nAcronym{IMDD}{\acroSCaps{im}\scslash \acroSCaps{dd}}{intensity-modulation direct-detection}
\nAcronym{IQM}{\acroSCaps{iqm}}{in-phase and quadrature modulator}
\nAcronym{ISI}{\acroSCaps{isi}}{inter-symbol interference}
\nAcronym{IP}{\acroSCaps{ip}}{intellectual property}

\nAcronym{JGN}{\acroSCaps{jgn}}{Japan Gigabit Network}

\nAcronym{KK}{\acroSCaps{kk}}{Kramers-Kronig}
\nAcronym{KIT}{\acroSCaps{kit}}{Karlsruhe Institute of Technology}

\nAcronym{LCOS}{\acroSCaps{LCoS}}{liquid crystal on silicon}
\nAcronym{LDPC}{\acroSCaps{ldpc}}{low-density parity-check}
\nAcronym{LEAF}{\acroSCaps{leaf}}{\usuk{large effective area fiber}{large effective area fibre}}
\nAcronym{LFSR}{\acroSCaps{lfsr}}{linear-feedback shift register}
\nAcronym{LG}{\acroSCaps{lg}}{Laguerre-Gaussian}
\nAcronym{LMS}{\acroSCaps{lms}}{least means squares}
\nAcronym{LLR}{\acroSCaps{llr}}{log-likelihood ratio}
\nAcronym{LO}{\acroSCaps{lo}}{local oscillator}
\nAcronym{LP}{\acroSCaps{lp}}{\usuk{linearly polarized}{linearly polarised}}
\nAcronym{LSPS}{\acroSCaps{lsps}}{\usuk{loop-synchronized polarization scrambler}{loop-synchronised polarisation scrambler}}
\nAcronym{LUT}{\acroSCaps{lut}}{lookup table}

\nAcronym{MVM}{\acroSCaps{mvm}}{matrix-vector multiplication}
\nAcronym{MB}{\acroSCaps{mb}}{Maxwell-Bolzmann}
\langcheck{%
    \nAcronym{MCF}{\acroSCaps{mcf}}{multi-core fiber}%
    }{%
    \nAcronym{MCF}{\acroSCaps{mcf}}{multi-core fibre}%
}%
\nAcronym{MDG}{\acroSCaps{mdg}}{mode dependent gain}
\nAcronym[firstplural=mode-dependent losses (\acroSCaps{mdl})]{MDL}{\acroSCaps{mdl}}{mode-dependent loss}
\nAcronym{MDM}{\acroSCaps{mdm}}{mode-division multiplexing}
\nAcronym{MEMS}{\acroSCaps{mems}}{micro-electro-mechanical systems}
\nAcronym{MF}{\acroSCaps{mf}}{matched filter}
\nAcronym{MFD}{\acroSCaps{mfd}}{mode field diameter}
\nAcronym{MI}{\acroSCaps{mi}}{mutual information}
\nAcronym{MIMO}{\acroSCaps{mimo}}{multiple-input multiple-output}
\nAcronym{ML}{\acroSCaps{ml}}{machine learning}
\nAcronym{MMA}{\acroSCaps{mma}}{multi-modulus algorithm}
\nAcronym{MMEDF}{\acroSCaps{mmedf}}{\usuk{multi-mode erbium-doped fiber}{multi-mode erbium-doped fibre}}
\nAcronym{MMEDFA}{\acroSCaps{mmedfa}}{\usuk{multi-mode erbium-doped fiber amplifier}{multi-mode erbium-doped fibre amplifier}}
\nAcronym{MMF}{\acroSCaps{mmf}}{\usuk{multi-mode fiber}{multi-mode fibre}}
\nAcronym{MMSE}{\acroSCaps{mmse}}{minimum mean squared error}
\nAcronym{MP}{\acroSCaps{mp}}{minimum phase}
\nAcronym{MPLC}{\acroSCaps{mplc}}{multi-plane light converter}
\nAcronym{MRC}{\acroSCaps{mrc}}{maximum ratio combining}
\nAcronym{MSE}{\acroSCaps{mse}}{mean squared error}
\nAcronym{MUX}{\acroSCaps{mux}}{multiplexer}
\nAcronym{MZM}{\acroSCaps{mzm}}{Mach-Zehnder modulator}
\nAcronym{MZI}{\acroSCaps{mzi}}{Mach-Zehnder interferometer}

\nAcronym{NA}{\acroSCaps{na}}{numerical aperture}
\langcheck{%
    \nAcronym{NANF}{\acroSCaps{nanf}}{nested antiresonant nodeless fiber}%
    }{%
    \nAcronym{NANF}{\acroSCaps{nanf}}{nested antiresonant nodeless fibre}%
}%
\nAcronym{NF}{\acroSCaps{nf}}{noise figure}
\nAcronym{NGMI}{\acroSCaps{ngmi}}{\usuk{normalized generalized mutual information}{normalised generalised mutual information}}
\nAcronym{NLSE}{\acroSCaps{nlse}}{nonlinear Schr\"{o}ding equation}
\nAcronym{NN}{\acroSCaps{nn}}{neural network}
\nAcronym{NIC}{\acroSCaps{nic}}{network interface card}
\nAcronym{NICT}{\acroSCaps{nict}}{National Institute of Information and Communications Technology}
\nAcronym{NIR}{\acroSCaps{nir}}{near-infrared}
\nAcronym{NISTSTS}{\acroSCaps{nist-sts}}{National Insitute of Standards and Technology: Statistical Test Suite}
\nAcronym{NRZ}{\acroSCaps{nrz}}{non-return-to-zero}
\nAcronym{NZDSF}{\acroSCaps{nz-dsf}}{\usuk{non-zero dispersion-shifted fiber}{non-zero dispersion-shifted fibre}} 

\nAcronym{OAM}{\acroSCaps{oam}}{orbital angular momentum}
\nAcronym{OBTB}{\acroSCaps{obtb}}{optical back-to-back}
\nAcronym{OCT}{\acroSCaps{oct}}{outer cladding thickness}
\nAcronym{ODE}{\acroSCaps{ode}}{ordinary differential equation}
\nAcronym{ODL}{\acroSCaps{odl}}{optical delay line}
\nAcronym{OEO}{\acroSCaps{oeo}}{optical-electrical-optical}
\nAcronym{OFC}{\acroSCaps{ofc}}{Optical Fiber Communications Conference}
\nAcronym{OFDR}{\acroSCaps{ofdr}}{optical frequency-domain reflectometer} 
\nAcronym{OFDM}{\acroSCaps{ofdm}}{orthogonal frequency division multiplexing}
\nAcronym{OH}{\acroSCaps{oh}}{overhead}
\nAcronym{OMFT}{\acroSCaps{omft}}{optical multi-format transmitter}
\nAcronym{OOK}{\acroSCaps{ook}}{on-off keying}
\nAcronym{OP}{\acroSCaps{op}}{optical processor}
\nAcronym{OPLL}{\acroSCaps{opll}}{optical phase-locked loop}
\nAcronym{OSA}{\acroSCaps{osa}}{\usuk{optical spectrum analyzer}{optical spectrum analyser}}
\nAcronym{OSNR}{\acroSCaps{osnr}}{optical signal-to-noise ratio}
\nAcronym{OTDR}{\acroSCaps{otdr}}{optical time-domain reflectometer}
\nAcronym{OTF}{\acroSCaps{otf}}{optical tunable filter}
\langcheck{%
    \nAcronym{OVNA}{\acroSCaps{ovna}}{optical vector network analyzer}%
    }{%
    \nAcronym{OVNA}{\acroSCaps{ovna}}{optical vector network analyser}%
}%
\nAcronym{OTG}{\acroSCaps{otg}}{optical turbulence generator}

\nAcronym{PAM}{\acroSCaps{pam}}{pulse-amplitude modulation}
\nAcronym{PAS}{\acroSCaps{pas}}{probabilistic amplitude shaping}
\nAcronym{PAPR}{\acroSCaps{papr}}{peak-to-average power ratio}
\nAcronym{PBC}{\acroSCaps{pbc}}{\usuk{polarization beam combiner}{polarisation beam combiner}}
\langcheck{%
    \nAcronym{PBGF}{\acroSCaps{pbgf}}{photonic bandgap fiber}%
    }{%
    \nAcronym{PBGF}{\acroSCaps{pbgf}}{photonic bandgap fibre}%
}%
\nAcronym{PBS}{\acroSCaps{pbs}}{polarization beam splitter}
\nAcronym{PC}{\acroSCaps{pc}}{physical contact}
\nAcronym{PolC}{\acroSCaps{PC}}{\usuk{polarization controller}{polarisation controller}}
\nAcronym{PCVD}{\acroSCaps{pcvd}}{plasma chemical vapor depostion}
\nAcronym{PCG}{\acroSCaps{pcg64}}{64-bit permuted congruential generator}
\nAcronym{PD}{\acroSCaps{pd}}{photodiode}
\nAcronym{PDF}{\acroSCaps{pdf}}{probability density function}
\langcheck{%
    \nAcronym{PDL}{\acroSCaps{pdl}}{polarization-dependent loss}
    }{%
    \nAcronym{PDL}{\acroSCaps{pdl}}{polarisation-dependent loss}
}%
\nAcronym{PDM}{\acroSCaps{pdm}}{\usuk{polarization-division multiplexing}{polarisation-division multiplexing}}
\langcheck{%
    \nAcronym{PER}{\acroSCaps{per}}{polarization extinction ratio}%
    }{%
    \nAcronym{PER}{\acroSCaps{per}}{polarisation extinction ratio}%
}%
\nAcronym{PIC}{\acroSCaps{pic}}{photonic integrated circuit}
\nAcronym{PL}{\acroSCaps{pl}}{photonic lantern}
\nAcronym{PMBPSK}{\acroSCaps{pm-bpsk}}{polarization-multiplexed binary phase-shift keying}
\nAcronym{PMQPSK}{\acroSCaps{pm-qpsk}}{polarization-multiplexed quaternary phase-shift keying}
\nAcronym{PM8QAM}{\acroSCaps{pm-8qam}}{polarization-multiplexed 8-ary quadrature amplitude modulation}
\nAcronym{PMD}{\acroSCaps{pmd}}{\usuk{polarization mode dispersion}{polarisation mode dispersion}}
\langcheck{%
    \nAcronym{PMF}{\acroSCaps{pmf}}{polarization-maintaining fiber}%
    }{%
    \nAcronym{PMF}{\acroSCaps{pmf}}{polarization-maintaining fibre}%
}%
\nAcronym{PMP}{\acroSCaps{pmp}}{phase-matching point}
\nAcronym{PNOB}{\acroSCaps{pnob}}{physical number of bits}
\nAcronym{PON}{\acroSCaps{pon}}{passive-optical network}
\nAcronym{PRBS}{\acroSCaps{prbs}}{pseudorandom bit sequence}
\nAcronym{PRNG}{\acroSCaps{PRNG}}{pseudo random number generator}
\nAcronym{PROFA}{\acroSCaps{profa}}{\usuk{pitch reducing optical fiber array}{pitch reducing optical fibre array}}
\nAcronym{PPM}{\acroSCaps{ppm}}{pulse-position modulation}
\nAcronym{PS}{\acroSCaps{ps}}{probabilistic shaping}
\langcheck{%
    \nAcronym{PSCF}{\acroSCaps{pscf}}{pure silica core fiber}%
    }{%
    \nAcronym{PSCF}{\acroSCaps{pscf}}{pure silica core fibre}%
}%
\nAcronym{PSD}{\acroSCaps{psd}}{power spectral density}
\nAcronym{PSF}{\acroSCaps{psf}}{point spread function}
\nAcronym{PSK}{\acroSCaps{psk}}{phase-shift keying}
\nAcronym{PSP}{\acroSCaps{psp}}{\usuk{principal states of polarization}{principal states of polarisation}}
\nAcronym{PSW}{\acroSCaps{psw}}{\usuk{polarization switch}{polarisation switch}}
\nAcronym{PID}{\acroSCaps{pid}}{proportional–integral–derivative}
\nAcronym{PV}{\acroSCaps{pv}}{process value}

\nAcronym{QAM}{\acroSCaps{qam}}{quadrature amplitude modulation}
\nAcronym{QKD}{\acroSCaps{qkd}}{quantum key distribution}
\nAcronym{QPSK}{\acroSCaps{qpsk}}{quaternary phase-shift keying}
\nAcronym{QRNG}{\acroSCaps{qrng}}{quantum random number generator}
\nAcronym{QSM}{\acroSCaps{qsm}}{quasi-single-mode}

\nAcronym{RAM}{\acroSCaps{ram}}{random-access memory}
\nAcronym{RC}{\acroSCaps{rc}}{raised cosine}
\nAcronym{RCMF}{\acroSCaps{rcmf}}{relative core multiplicity factor}
\nAcronym{RF}{\acroSCaps{rf}}{radio frequency}
\nAcronym{RFSoC}{\acroSCaps{rfsoc}}{radio frequency system-on-chip}
\nAcronym{RI}{\acroSCaps{ri}}{refractive index}
\nAcronym{RLS}{\acroSCaps{rls}}{recursive least squares}
\nAcronym{RRC}{\acroSCaps{rrc}}{root-raised-cosine}
\nAcronym{ROADM}{\acroSCaps{roadm}}{reconfigurable optical add-drop multiplexer}
\nAcronym{ROI}{\acroSCaps{roi}}{region of interest} 
\nAcronym{RTO}{\acroSCaps{rto}}{real-time oscilloscope} 
\nAcronym{RZDBPSK}{\acroSCaps{rz-dbpsk}}{return-to-zero differential binary phase-shift keying} 
\nAcronym{RZDQPSK}{\acroSCaps{rz-dqpsk}}{return-to-zero differential quaternary phase-shift keying}
\nAcronym{RN}{\acroSCaps{RN}}{random number}

\nAcronym{S2}{\acroSCaps{S\textsuperscript{2}}}{spatially and spectrally resolved}
\nAcronym{SA}{\acroSCaps{sa}}{simulated annealing}
\nAcronym{SamPerSym}{\acroSCaps{sps}}{samples per symbol}
\nAcronym{SBS}{\acroSCaps{sbs}}{stimulated Brillouin scattering}
\nAcronym{SCM}{\acroSCaps{scm}}{subcarrier multiplexing}
\nAcronym{SDFEC}{\acroSCaps{sd-fec}}{soft-decision forward error correction}
\nAcronym{SDM}{\acroSCaps{sdm}}{space-division multiplexing}
\nAcronym{SE}{\acroSCaps{se}}{spectral efficiency}
\nAcronym{SER}{\acroSCaps{ser}}{symbol error rate}
\nAcronym{SI}{\acroSCaps{si}}{step index}
\nAcronym{SIFMF}{\acroSCaps{si-fmf}}{\usuk{step-index few-mode fiber}{step-index few-mode fibre}}
\nAcronym{SISMF}{\acroSCaps{si-smf}}{\usuk{step-index single-mode fiber}{step-index single-mode fibre}}
\nAcronym{SLM}{\acroSCaps{slm}}{spatial light modulator}
\nAcronym{SKR}{\acroSCaps{skr}}{secret key rate}
\nAcronym{SMD}{\acroSCaps{smd}}{spatial-mode dispersion}
\nAcronym{SMF}{\acroSCaps{smf}}{\usuk{single-mode fiber}{single-mode fibre}}
\nAcronym{SMUX}{\acroSCaps{smux}}{spatial multiplexer}
\nAcronym{SNR}{\acroSCaps{snr}}{signal-to-noise ratio}
\nAcronym{SNU}{\acroSCaps{snu}}{shot-noise unit}
\nAcronym{SOA}{\acroSCaps{soa}}{semiconductor optical amplifier}
\nAcronym[firstplural=\usuk{states of polarization (\acroSCaps{sop})}{states of polarisation (\acroSCaps{sop})}]{SOP}{\acroSCaps{sop}}{\usuk{state of polarization}{state of polarisation}}
\nAcronym{SPM}{\acroSCaps{spm}}{self-phase modulation}
\nAcronym{SPS}{\acroSCaps{sps}}{samples per symbol}
\nAcronym{SRS}{\acroSCaps{srs}}{stimulated Raman scattering}
\nAcronym{SSB}{\acroSCaps{ssb}}{single-sideband}
\nAcronym{SSBI}{\acroSCaps{ssbi}}{signal-signal beat interference}
\nAcronym{SSFM}{\acroSCaps{ssfm}}{split-step Fourier method}
\nAcronym{SSMF}{\acroSCaps{ssmf}}{\usuk{standard single-mode fiber}{standard single-mode fibre}}
\nAcronym{STAXT}{\acroSCaps{staxt}}{short-term average cross-talk}
\nAcronym{STL}{\acroSCaps{stl}}{swept tunable laser}
\nAcronym{SVD}{\acroSCaps{svd}}{singular value decomposition}
\nAcronym{SW}{\acroSCaps{sw}}{sequence-wise}
\nAcronym{SWI}{\acroSCaps{swi}}{swept wavelength interferometry}
\nAcronym{SP}{\acroSCaps{sp}}{setpoint}

\nAcronym{TAT}{\acroSCaps{tat}}{transatlantic}
\nAcronym{TD}{\acroSCaps{td}}{time domain}
\nAcronym{TDE}{\acroSCaps{tde}}{\usuk{time domain equalizer}{time domain equaliser}}
\nAcronym{TDFA}{\acroSCaps{tdfa}}{thulium doped-fiber amplifier}
\nAcronym{TDM}{\acroSCaps{tdm}}{time-domain multiplexing}
\nAcronym{TDMSDM}{\acroSCaps{tdm-sdm}}{time-domain multiplexed space-division multiplexing}
\nAcronym{TE}{\acroSCaps{te}}{transverse electric}
\nAcronym{TRNG}{\acroSCaps{TRNG}}{true random number generator}
\nAcronym{TEC}{\acroSCaps{tec}}{thermally-expanded-core}
\nAcronym{TLS}{\acroSCaps{tls}}{tunable laser source}
\nAcronym{TIA}{\acroSCaps{tia}}{trans-impedance amplifier}
\nAcronym{TM}{\acroSCaps{tm}}{transverse magnetic}
\nAcronym{TH4D}{\acroSCaps{th-4d}}{time domain hybrid four-dimensional}
\nAcronym{TH4D2A8PSK}{\acroSCaps{th-4d-2a8psk}}{time-domain hybrid four-dimensional two-amplitude eight-phase-shift keying}
\nAcronym{TP}{\acroSCaps{tp}}{twisted pair}
\nAcronym{TTL}{\acroSCaps{ttl}}{transistor-transistor logic}

\nAcronym{UWB}{\acroSCaps{uwb}}{ultra-wideband}
\nAcronym{ULI}{\acroSCaps{uli}}{ultrafast laser inscription}

\nAcronym{VCSEL}{\acroSCaps{vcsel}}{vertical-cavity surface emitting laser}
\nAcronym{VHDL}{\acroSCaps{vhdl}}{\acroSCaps{Vhsic} Hardware Description Language}
\nAcronym{VOA}{\acroSCaps{voa}}{variable optical attenuator}

\nAcronym{WDL}{\acroSCaps{wdl}}{wavelength-dependent loss}
\nAcronym{WDM}{\acroSCaps{wdm}}{wavelength-division multiplexing}
\nAcronym{WFS}{\acroSCaps{wfs}}{wavefront sensor}
\nAcronym{WGA}{\acroSCaps{wga}}{weakly guiding approximation}
\nAcronym{WGN}{\acroSCaps{wgn}}{white Gaussian noise}
\nAcronym[longplural=wavelength selective switches]{WSS}{\acroSCaps{wss}}{wavelength selective switch}
\nAcronym{WC}{\acroSCaps{wc}}{weakly coupled}
\nAcronym{WCMCF}{\acroSCaps{wc-mcf}}{\usuk{weakly-coupled multi-core fiber}{weakly-coupled multi-core fibre}}

\nAcronym{XGM}{\acroSCaps{xgm}}{cross-gain modulation}
\nAcronym{XPM}{\acroSCaps{xpm}}{cross-phase modulation}
\nAcronym{XOR}{\acroSCaps{xor}}{exclusive or}
\langcheck{%
    \nAcronym{XPOLM}{\acroSCaps{xp}ol\acroSCaps{m}}{cross-polarization modulation}
    }{%
    \nAcronym{XPOLM}{\acroSCaps{xp}ol\acroSCaps{m}}{cross-polarisation modulation}
}%
\nAcronym{XT}{\acroSCaps{xt}}{cross-talk}

\nAcronym{3DWG}{\acroSCaps{3dwg}}{3D-waveguide}